\documentclass[preprint,12pt]{aastex}
\usepackage{amsmath}

\newcommand{\eqw}{\ensuremath{{\mathcal W}_{\lambda}}}
\newcommand{\ewlim}{\ensuremath{{\mathcal W}_{\rm lim}}}
\newcommand{\hst}{{\sl HST}}
\newcommand{\kms}{\ensuremath{{\rm km\,s}^{-1}}}

\begin{document}

\title{On the Significance of Absorption Features in \hst/COS Data\altaffilmark{*}}
\altaffiltext{*}{Based on observations with the NASA/ESA {\sl Hubble Space Telescope}, obtained at the Space Telescope Science Institute, which is operated by the Associated Universities for Research in Astronomy, Inc., under NASA contract NAS 5-26555.}

\author{Brian A. Keeney, Charles W. Danforth, John T. Stocke, Kevin France, \& James C. Green}
\affil{Center for Astrophysics and Space Astronomy, Department of Astrophysical and Planetary Sciences, University of Colorado, \\ 389 UCB, Boulder, CO 80309, USA; brian.keeney@colorado.edu}

\shorttitle{On the Significance of COS Absorption Lines}
\shortauthors{Keeney et~al.}

\begin{abstract}
We present empirical scaling relations for the significance of absorption features detected in medium resolution, far-UV spectra obtained with the Cosmic Origins Spectrograph (COS).  These relations properly account for both the extended wings of the COS line spread function and the non-Poissonian noise properties of the data, which we characterize for the first time, and predict limiting equivalent widths that deviate from the empirical behavior by $\leq5$\% when the wavelength and Doppler parameter are in the ranges $\lambda=1150$--1750~\AA\ and $b>10$~\kms.  We have tested a number of coaddition algorithms and find the noise properties of individual exposures to be closer to the Poissonian ideal than coadded data in all cases.  For unresolved absorption lines, limiting equivalent widths for coadded data are 6\% larger than limiting equivalent widths derived from individual exposures with the same signal-to-noise.  This ratio scales with $b$-value for resolved absorption lines, with coadded data having a limiting equivalent width that is 25\% larger than individual exposures when $b\approx150$~\kms.
\end{abstract}


\section{Introduction}
\label{intro}

The Cosmic Origins Spectrograph (COS) was installed aboard the {\sl Hubble Space Telescope} (\hst) in May 2009 as part of \hst\ Servicing Mission 4.  COS is a slitless spectrograph with far-UV (FUV) and near-UV spectroscopic and imaging modes.  Its design philosophy and on-orbit performance are detailed in \citet{green12} and \citet{osterman11}.

 COS has three FUV gratings: the medium resolution G130M and G160M gratings, which use five central wavelength settings to cover the wavelength ranges 1135--1470~\AA\ and 1385--1795~\AA, respectively, at a spectroscopic resolving power of $R \equiv \lambda / \Delta \lambda \approx 18000$; and the low resolution G140L grating, which uses two central wavelength settings to cover the wavelength range 900--2150~\AA\ at a resolving power of $R \approx 2500$ \citep{dixon11,green12}.  Three additional central wavelengths for the G130M grating have become available recently \citep{dixon11}, but since they are not as well-characterized as the original G130M settings we do not investigate them here.  All of the gratings disperse light, re-image the diverging telescope beam, and correct for the astigmatism of the spectrograph design and the spherical aberration of the \hst\ primary mirror in a single optical element \citep{osterman11,green12}. This design minimizes reflective losses in the system, creating a spectrograph with unprecedented FUV sensitivity; the COS FUV throughput is 10--20 times greater than comparable modes of previous \hst\ FUV spectrographs \citep{green12}.

Thermal vacuum testing of the COS resolving power pre-launch found that the line spread function (LSF) of the G130M and G160M gratings was well-approximated by a Gaussian with a FWHM of $\sim\,$6.5~pixels ($\Delta v \sim 17$~\kms).  However, the measured on-orbit LSF of these (and all other) gratings was found to have extended non-Gaussian wings that vary as a function of wavelength. These wings are caused by mid-frequency wavefront errors in the \hst\ primary mirror \citep{ghavamian09} and are further enhanced by a small but measurable amount of scattering caused by microroughness on the primary mirror surface \citep{kriss11}.  These deviations from a Gaussian profile affect COS more than \hst's previous UV spectrographs because of its slitless design, and reduce the instrument's sensitivity to weak absorption features by decreasing the resolution and intensity at the core of the LSF.

In this Paper, we expand upon the work of \citet{ghavamian09} and \citet{kriss11} by generalizing their equivalent width formalism to spectroscopically resolved lines and investigating the effects of non-Poissonian noise in COS G130M and G160M data on the significance of absorption features.  While important for all scientific applications, this non-Poissonian noise affects the detectability of broad, shallow absorption features in particular, such as Ly$\alpha$ and metal-line absorption associated with the warm-hot intergalactic medium \citep*{savage10,savage11,narayanan11,danforth10b,danforth11}.  In Section~\ref{noise} we demonstrate the presence of non-Poissonian noise in COS FUV data and characterize its effects on the limiting equivalent widths of absorption lines.  In Section~\ref{scaling} we derive empirical approximations for the significance of absorption features detected in COS FUV data that properly account for both the extended wings of the on-orbit COS LSF and the non-Poissonian noise properties of the data.  These approximations predict equivalent widths that differ from the empirical behavior by $\leq5$\% over the entire G130M + G160M bandpass (1150--1750~\AA) when the Doppler parameter $b>10$~\kms.  Finally, we summarize our results and most important conclusions in Section~\ref{conclusion}.

\section{Non-Poissonian Noise Properties}
\label{noise}

When examining a spectrum, one is often interested in knowing the smallest equivalent width that can be measured to a certain significance level.  This limiting equivalent width was expressed by \citet{ghavamian09} as
\begin{equation}
\ewlim = \frac{N_{\sigma} \, \lambda_x}{{\rm (S/N)}_x} \, \frac{1}{f_c(x)},
\label{eqn:ewlim}
\end{equation}
where $x$ is the width (in pixels) of the discrete region of integration over which the limit is calculated, $N_{\sigma}$ is the significance level of the limit expressed as a multiple of $\sigma$ for Gaussian distributed errors, $\lambda_x = \Delta\lambda \times x$ is the width of the integration region in wavelength space, $\Delta\lambda$ is the dispersion of the spectrum \citep[9.97~${\rm m\mbox{\AA}\,pixel^{-1}}$ for the G130M grating and 12.23~${\rm m\mbox{\AA}\,pixel^{-1}}$ for G160M;][]{dixon11}, ${\rm (S/N)}_x$ is the signal-to-noise ratio of the spectrum at a resolution of $\lambda_x$, and $f_c(x)$ is the fractional area of the LSF contained within the region of integration.

If the noise in the spectrum is purely Poissonian, ${\rm (S/N)}_x = {\rm (S/N)}_1 \times x^{1/2}$, where ${\rm (S/N)}_1$ is the signal-to-noise per pixel.  \citet{ghavamian09} use this relation to derive the limiting equivalent width to be
\begin{equation}
\ewlim = \frac{N_{\sigma} \, \Delta\lambda}{{\rm (S/N)}_1} \; \frac{x^{1/2}}{f_c(x)}.
\label{eqn:ewlim2}
\end{equation}
The limiting equivalent width is obviously a function of $x$, the size of the region of integration.  Fortunately, $x^{1/2}/f_c(x)$ can be uniquely minimized such that there is an optimum integration width, $x_{\rm opt}$, that provides the most stringent equivalent width limit.  Note that for COS data  $f_c$ is a function of both $x$ and $\lambda$ since the wings of the LSF vary with wavelength \citep{ghavamian09,kriss11}.  Thus, $x_{\rm opt}$ and \ewlim\ are wavelength dependent even for fixed signal-to-noise.

However, the noise in COS FUV data is not purely Poissonian.  We define the measured relationship between the smoothed signal-to-noise and the signal-to-noise per pixel to be
\begin{equation}
\eta(x) \equiv \frac{{\rm (S/N)}_x}{{\rm (S/N)}_1},
\label{eqn:eta}
\end{equation}
where $\eta(x) = x^{1/2}$ for purely Poissonian noise.  To quantify the functional form of $\eta(x)$ we have measured ${\rm (S/N)}_x$ as a function of $x$ at several wavelengths for all of the FUV data obtained as part of COS GTO programs 11520 and 12025 (G130M + G160M spectra of 16~QSOs with coadded ${\rm (S/N)}_1 \approx 5$--25).  All exposures were reduced with \textsc{CalCOS v2.17.3a} and processed with custom IDL routines for aligning and coadding COS FUV data as detailed in \citet{danforth10a} and \citet{keeney12}.  The signal-to-noise ratio was calculated by identifying line-free continuum regions, smoothing the spectra by $x$ pixels, and measuring the rms continuum deviations in the smoothed spectra.  

\begin{figure}
\epsscale{1.00}
\centering \plotone{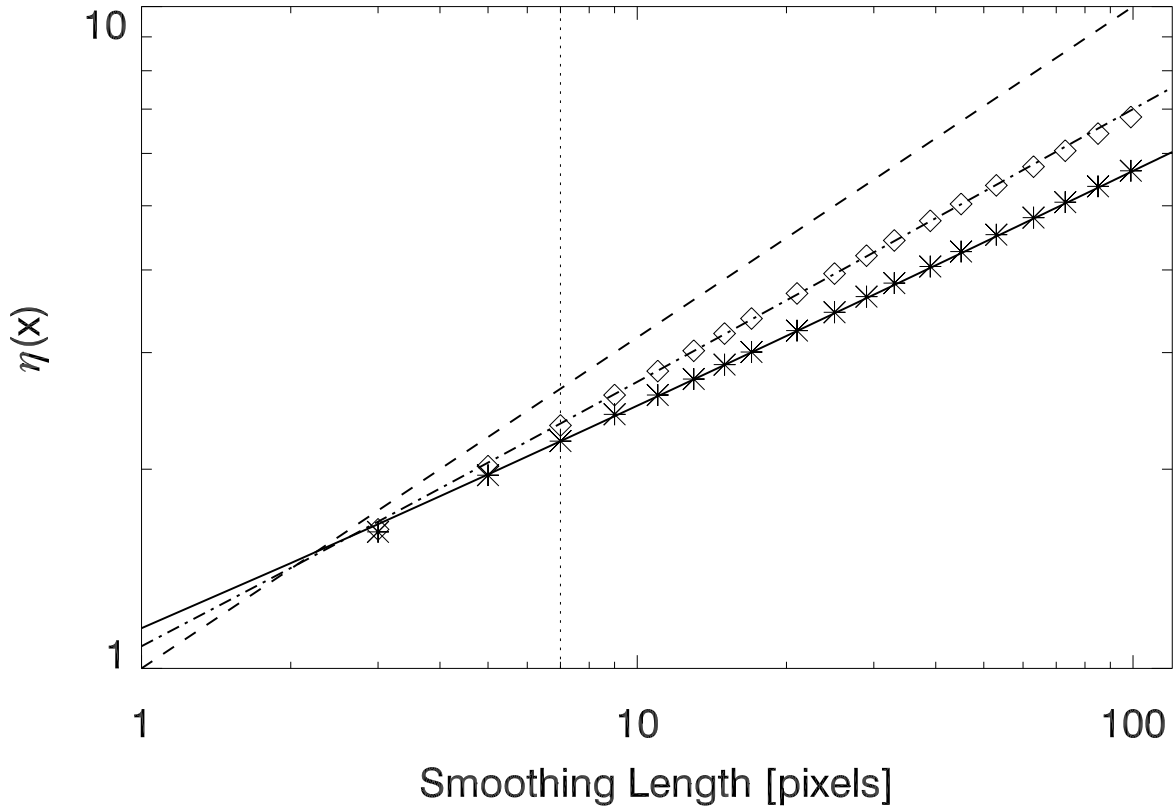}
\caption{The ratio of the smoothed signal-to-noise to the signal-to-noise per pixel, $\eta(x)$, as a function of smoothing length, $x$.  For purely Poissonian noise, $\eta(x) \equiv x^{1/2}$ (dashed line).  We find $\eta(x) \approx x^{0.42} + 0.08$ (dot-dashed line) for individual exposures and $\eta(x) \approx x^{0.37} + 0.15$ (solid line) for coadded data (see Table~\ref{tab:coeff}), indicating the presence of non-Poissonian noise.  Only smoothing lengths larger than the FWHM of the COS on-orbit LSF (dotted vertical line; see Table~\ref{tab:ewlim}) contribute to the fits.
\label{fig:eta}}
\end{figure}

Figure~\ref{fig:eta} shows our measured values of $\eta(x)$ as a function of $x$ for both individual exposures (diamonds) and coadded data (asterisks).  The plotted symbols represent the average value of $\eta$ as determined from our 16~sight lines, each of which had measurements taken at 13 different wavelengths.  We find that $\eta$ is well fit by a power law of the form
\begin{equation}
\eta(x) = \eta_0 + x^{\beta}.
\label{eqn:eta2}
\end{equation}
The solid line in Figure~\ref{fig:eta} indicates our best-fit power law for coadded data, and the dot-dashed line indicates our best-fit power law for individual exposures.  The dashed line illustrates the behavior of Poissonian noise.  Only smoothing lengths larger than the FWHM of the on-orbit LSF (dotted vertical line; see Table~\ref{tab:ewlim}) contribute to the fits.  Our best-fit values of $\eta_0$ and $\beta$ for Poissonian data, individual exposures, and coadded data are listed in Table~\ref{tab:coeff}.

We find a significant difference in $\eta(x)$ measured from individual exposures and coadded data, but emphasize that even though the non-Poissonian noise properties of coadded data are more severe than for individual exposures, the coadded signal-to-noise is significantly higher, allowing observers to set more stringent equivalent width limits from coadded data in most cases (see discussion in Section~\ref{scaling:coadd}).  We find no evidence for systematic trends in $\eta(x)$ as a function of wavelength or signal-to-noise in either individual exposures or coadded data.  For coadded data, we also find no evidence for systematic trends in $\eta(x)$ as a function of the number of exposures being added together.   

The G140L grating is not the best choice for detecting weak spectral features, so we do not describe it in detail here.  However, we note that \citet{syphers12} find $\eta \simeq x^{0.44}$ for small smoothing lengths ($x\lesssim14$~pixels), and $\eta \simeq x^{0.42}$ for larger smoothing lengths ($x\lesssim40$~pixels) in individual G140L exposures.  These values are both close to our measured value of $\eta \approx 0.08 + x^{0.42}$ for individual G130M and G160M exposures (see Table~\ref{tab:coeff}), suggesting that the non-Poissonian noise in individual COS exposures originates in the FUV detector itself rather than any particular grating.

The COS FUV detector utilizes two microchannel plate (MCP) segments, which both have dead spots, gain variations, and shadows from the ion-repeller grid wires used to improve detector quantum efficiency. Additionally, each segment has irregularities introduced by hexagonal and moir\'e patterns in the MCPs themselves.  All of these features create fixed-pattern (i.e., non-Poissonian) noise in COS FUV spectra when these spatially variable two-dimensional ``pixels'' are compressed to one dimension as part of the spectral extraction.  While some of these effects (e.g., grid wire shadows) can be corrected by flat fielding, they cannot be removed altogether \citep{ake10}.

Standard observing strategy suggests that targets should be observed using multiple FP-POS positions and/or central wavelength settings to reduce the effects of fixed-pattern noise \citep[see Section~5.8.2 of][]{dixon11}, but we find just the opposite effect.  Figure~\ref{fig:eta} shows that $\eta(x)$ for coadded data is further from the Poissonian expectation than $\eta(x)$ for individual exposures.  This result is not in itself surprising given the origin of the non-Poissonian noise in individual COS exposures and the spectrograph's slitless design, since the hexagonal and moir\'e patterns of the MCPs can create large fixed-pattern differences from small variations in the spectrum's position in the detector's cross-dispersion direction.  In general, these differences do not simply average out, and can potentially constructively interfere to create amplified fixed-pattern noise; indeed, this is one of the reasons why a robust one-dimensional FUV flat field \citep{ake10} has been difficult to produce for COS.

If the signal-to-noise of a dataset is high enough (${\rm S/N} \gtrsim 10$ per pixel in individual exposures) an iterative process can be utilized, which was first developed for high signal-to-noise observations obtained with the Goddard High Resolution Spectrograph \citep{cardelli93,lambert94}.  This iterative procedure can be used to create a custom flat field that improves the final signal-to-noise of individual exposures and coadded data for that specific dataset \citep{ake10}.  Given the higher signal-to-noise achieved in data processed by these custom flat fields compared to the standard flat fields applied by \textsc{CalCOS}, the non-Poissonian behavior should be less severe when a custom flat field is employed.  The Mrk~421 sight line studied by \citet{danforth11} is the only one in our sample with sufficient signal-to-noise to utilize this technique, and we note a modest improvement in the noise properties of this dataset when a custom flat field is employed \citep[as was done in][]{danforth11} rather than the standard \textsc{CalCOS} flat field.

Surprisingly, we find that the severity of the non-Poissonian noise in coadded COS data depends on the coaddition algorithm employed.  In particular, we have found that the process of interpolating individual COS exposures onto a new wavelength scale (a necessary step in coadding multiple exposures with only partially overlapping wavelength ranges) can introduce additional non-Poissonian noise to the coaddition process.  We have tested several algorithms, including linear, quadratic, cubic spline, and nearest neighbor interpolation schemes, and found that data coadded using nearest neighbor interpolation has the least severe non-Poissonian noise.  The coadded data points in Figure~\ref{fig:eta}, for which we found $\eta(x) \propto x^{0.37}$ (see Table~\ref{tab:coeff}), utilize nearest neighbor interpolation. We find $\eta \propto x^{0.33}$ for data coadded using linear interpolation\footnote{Our coaddition routine originally used a linear interpolation scheme, but has been modified to use nearest neighbor interpolation instead as a result of this testing.  Consequently, we recommend that users who downloaded previous versions of our coaddition code upgrade to the latest version (v2.0) at \url{http://casa.colorado.edu/$\sim$danforth/costools.html}.} and $\eta \propto x^{0.36}$ for data coadded using quadaratic or cubic spline interpolation.  We also find that the ``x1dsum'' files generated by \textsc{CalCOS} have $\eta \propto x^{0.33}$, identical to the behavior we find for linear interpolation.

Regardless of its origin, non-Poissonian noise has an appreciable effect on the limiting equivalent width achievable with COS data.  In the following subsections we calculate equivalent width limits for unresolved (Section~\ref{noise:unresolved}) and spectroscopically resolved (Section~\ref{noise:resolved}) absorption lines that account for the non-Poissonian noise in the data, and compare them to the limits derived by \citet{ghavamian09} and \citet{kriss11}, which assume purely Poissonian noise.

\subsection{Unresolved Lines}
\label{noise:unresolved}

If we substitute the results of Equation~\ref{eqn:eta} into Equation~\ref{eqn:ewlim}, we find
\begin{equation}
\ewlim = \frac{N_{\sigma} \, \Delta\lambda}{{\rm (S/N)}_1} \; \frac{x}{\eta(x) \, f_c(x)} = \frac{N_{\sigma} \, \Delta\lambda}{{\rm (S/N)}_1} \; w(x),
\label{eqn:ewlim3}
\end{equation}
which reduces to Equation~\ref{eqn:ewlim2} in the Poissonian limit.  The function $w(x)$ has a unique minimum for all values of $0 \leq \beta \leq 1$ (see Eqn.~\ref{eqn:eta2}) if $f_c$ is monotonic with $x$, as is the case for any reasonable LSF.  This implies that a single value, $x_{\rm opt}$, exists that sets the most stringent limit on \ewlim.  Physically, $x_{\rm opt}$ represents the largest number of pixels that can be summed over a line profile before the noise contributions begin to outweigh the flux contributions \citep{ghavamian09}.

\begin{figure}
\epsscale{1.00}
\centering \plotone{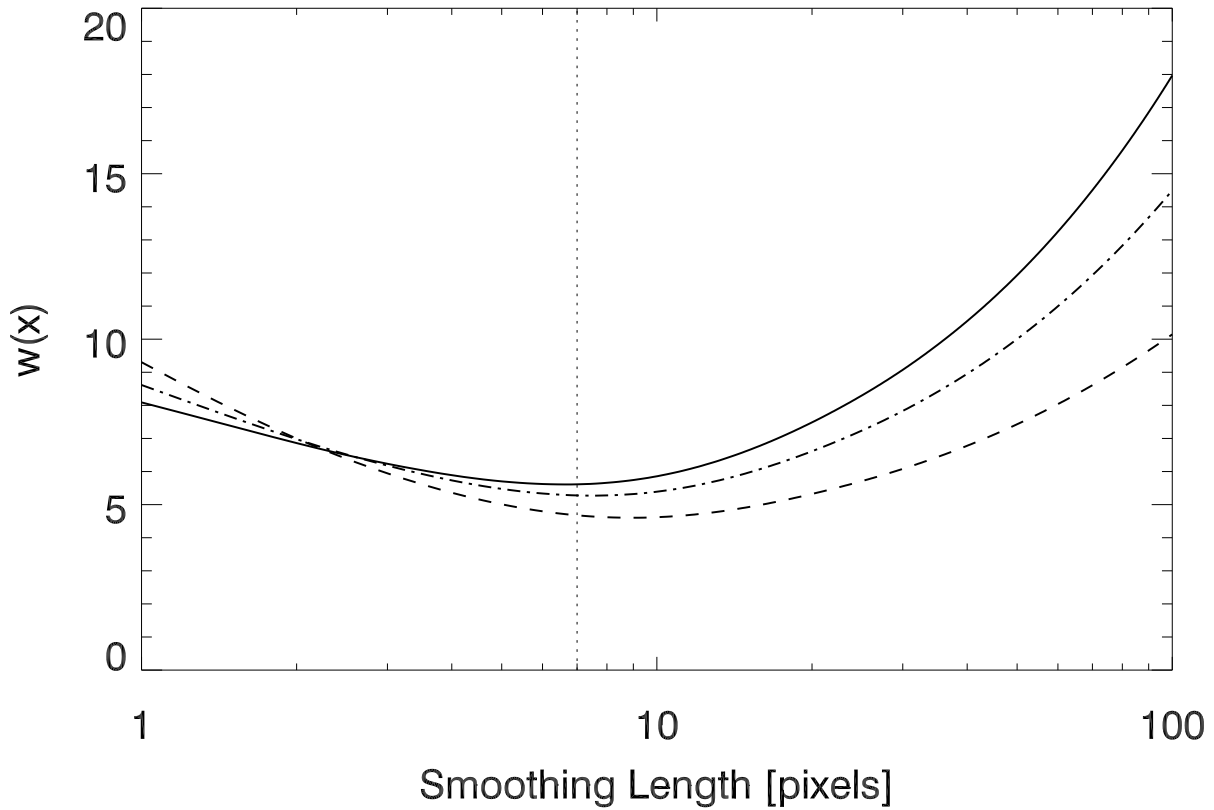}
\caption{The function $w(x)$ (see Eqn.~\ref{eqn:ewlim3}) evaluated at $\lambda=1250$~\AA\ for coadded data (solid line) and individual exposures (dot-dashed line), compared to the Poissonian value (dashed line).  Non-Poissonian noise leads to a smaller value of $x_{\rm opt}$ and a larger minimum for the function.  The dotted vertical line shows the approximate FWHM of the COS on-orbit LSF, which is approximately equal to $x_{\rm opt}$ for individual exposures (see Table~\ref{tab:ewlim}).
\label{fig:min}}
\end{figure}

Figure~\ref{fig:min} shows how $w(x)$ varies with $x$ for coadded data (solid line) and individual exposures (dot-dashed line), as compared to Poissonian data (dashed line) at a wavelength of 1250~\AA.  Non-Poissonian noise causes $x_{\rm opt}$ to be smaller and the minimum value of $w(x)$ to be larger for coadded data and individual exposures than in the Poissonian case.  Both of these effects are caused by the measured ${\rm (S/N)}_x$ being smaller than Poisson statistics predict.  

Table~\ref{tab:ewlim} illustrates how non-Poissonian noise affects the limiting equivalent width of unresolved absorption lines in COS G130M and G160M spectra.  We list the FWHM, $x_{\rm opt}$, and limiting $3\sigma$ equivalent width of the \citet{ghavamian09} and \citet{kriss11} LSFs as a function of wavelength for the G130M and G160M gratings, as well as the corresponding values for the best-fit Gaussian LSF from pre-launch thermal vacuum testing.  For each LSF, we tabulate these quantities assuming purely Poissonian noise for comparison with the values in Table~1 of \citet{ghavamian09} and Table~3 of \citet{kriss11}, which we are able to reproduce.  There is a negligible difference in FWHM and $x_{\rm opt}$ between the two LSFs but the \citet{kriss11} LSF produces equivalent width limits that are $\sim2$\% larger than those derived from the \citet{ghavamian09} LSF.

For the \citet{kriss11} LSF we also tabulate the FWHM, $x_{\rm opt}$, and limiting $3\sigma$ equivalent width values for individual COS exposures and coadded COS data.  These values allow us to quantify the trends that we illustrated in Figure~\ref{fig:min}:  for unresolved lines, $x_{\rm opt}$ is $\sim15$\% smaller and $\ewlim(3\sigma)$ is $\sim15$\% larger for individual exposures than Poissonian data, and  $x_{\rm opt}$ is $\sim25$\% smaller and $\ewlim(3\sigma)$ is $\sim20$\% larger for coadded data than Poissonian data.  These trends hold for the Gaussian LSF as well but their magnitude is slightly smaller due to the absence of the extended wings of the measured COS on-orbit LSF.  Finally, we note that $x_{\rm opt}$ for individual exposures is always $\simeq0.2$~pixels larger than the FWHM of the LSF, and $x_{\rm opt}$ for coadded data is always $\simeq0.6$~pixels smaller than the FWHM of the LSF (i.e., $x_{\rm opt} \simeq 7.4$~pixels for individual exposures and $x_{\rm opt} \simeq 6.5$~pixels for coadded data; see Table~\ref{tab:ewlim}).

\subsection{Spectroscopically Resolved Lines}
\label{noise:resolved}

The equivalent width formalism of Equation~\ref{eqn:ewlim3} can be extended to spectroscopically resolved absorption lines by modifying the fractional area of the line profile contained within the region of integration, $f_c(x)$.  While for unresolved lines the line profile chosen was simply the instrumental LSF, the line profile of resolved features is the convolution of the LSF with the intrinsic shape of the absorption line.  Generically, the intrinsic absorption line profile is described by a Voigt function that may have multiple velocity components; however, since we are only considering our ability to detect very weak features when calculating \ewlim\ it is appropriate to assume a single-component Gaussian line profile.

Convolving the instrumental LSF with the intrinsic line shape introduces a generic dependence on the Doppler parameter, $b$, to $f_c$ (i.e., $f_c = f_c(x,b)$).  One consequence of this dependence is that calculating \ewlim\ requires assuming a $b$-value for the absorption line of interest.  Determining an appropriate $b$-value is most straightforward if one is setting a limit on a line of the same ionic species as one of the lines detected in the spectrum (e.g., if one is trying to set a limit on the weaker line of a doublet when the stronger line is detected).  Furthermore, as discussed in Section~\ref{noise} and \citet{ghavamian09}, the wings of the COS LSF are wavelength dependent, introducing a wavelength dependence to $f_c$ for COS data such that $f_c = f_c(x,\lambda,b)$.

Table~\ref{tab:resolved} lists the limiting equivalent width of resolved absorption features in COS G130M and G160M spectra assuming a Gaussian line profile with Doppler parameter, $b$. This Gaussian profile was convolved with the \citet{kriss11} LSF and we have tabulated the Gaussian and convolved FWHMs, as well as $x_{\rm opt}$, $f_c(x_{\rm opt})$ and $\ewlim(3\sigma)$ for Poissonian data, individual exposures, and coadded data.  For sufficiently large $b$-values ($\gtrsim\,$20~\kms) the convolved FWHM is always $\sim\,$2--3~pixels larger than the Gaussian FWHM; for smaller $b$-values the line is only marginally resolved and the convolution has a larger effect.  Thus, the convolution acts to increase the apparent $b$-value of a line by approximately $1.5$--2~pixels ($\sim4$~\kms\ at 1450~\AA; see Eq.~\ref{eqn:dx} and \ref{eqn:fwhm}) if the effects of this convolution are not accounted for.

The effect of non-Poissonian noise on $x_{\rm opt}$ for resolved lines is nearly identical to its effect on $x_{\rm opt}$ for unresolved lines (see Section~\ref{noise:unresolved}, Table~\ref{tab:ewlim}), but its magnitude decreases by $\sim3$\% between $b=10$~\kms\ and $b=200$~\kms\ as the shape of the convolved line profile becomes more and more Gaussian.  The effect on $\ewlim(3\sigma)$ is striking, however, with the limits for individual exposures increasing from 120\% to 150\% of the Poissonian values, and the limits for coadded data increasing from 130\% to 190\% of the Poissonian values, as the Doppler parameter increases from $b=10$~\kms\ to $b=200$~\kms.

\section{Empirical Scaling Relations}
\label{scaling}

We are now interested in determining the significance level of an absorption feature detected by the COS medium resolution gratings.  We can do so by manipulating Equations~\ref{eqn:ewlim} and \ref{eqn:ewlim3}, replacing the limiting equivalent widths (\ewlim) defined therein with an arbitrary equivalent width, \eqw:
\begin{align}
\label{eqn:siglevel}
N_{\sigma}(x,\lambda,b) &= {\rm (S/N)}_x \, \frac{\eqw}{\Delta\lambda} \; \frac{f_c(x,\lambda,b)}{x} \\ 
                        &= {\rm (S/N)}_1 \, \frac{\eqw}{\Delta\lambda} \; \frac{\eta(x)}{x} \; f_c(x,\lambda,b) \nonumber \\
                        &= {\rm (S/N)}_1 \, \frac{\eqw}{\Delta\lambda} \; \frac{1}{w(x,\lambda,b)}, \nonumber
\end{align}
where we have now explicitly listed the dependence of $f_c$ on wavelength and Doppler parameter and $w$ is defined as in Equation~\ref{eqn:ewlim3}.  While Equation~\ref{eqn:siglevel} is algebraically simple, $f_c$ depends on the details of the LSF, making the equation cumbersome to use in practice.  In particular, for sufficiently large \eqw, our assumption of an intrinsically Gaussian line shape with a single velocity component (see Section~\ref{noise:resolved}) breaks down and there is no way of determining a unique value of $f_c$.  However, it has been our experience that observers are most interested in determining whether an absorption feature is detected above a modest given significance threshold ($3\sigma$, $4\sigma$, etc.); for lines that are clearly detected above this threshold (i.e., where our assumption about the intrinsic line shape typically breaks down), accurately determining the exact value of the significance level of the detection is generally not important.

To maximize the significance level of an observed line, one should integrate over a discrete region $x_{\rm opt}$ pixels wide and evaluate $w$ at $x=x_{\rm opt}$.  In \citet{danforth11} we made the simplifying assumption that the absorption profile's deviations from Gaussianity were small, such that $x_{\rm opt}$ equals the Gaussian FWHM and $f_c(x_{\rm opt}) \approx 0.761$, to reduce Equation~\ref{eqn:siglevel} to a more tractable form \citep[see Eq.~4 of][]{danforth11}.  For coadded data, this assumption is justified if $b \gtrsim 30$~\kms, where $x_{\rm opt}$ and $f_c(x_{\rm opt})$ both differ from their assumed values by $<10$\% (see Table~\ref{tab:resolved}).

Here we search for parameterizations of $x_{\rm opt}$ and $f_c(x_{\rm opt})$ that are valid for a broader range of $b$-values, allowing us to reduce Equation~\ref{eqn:siglevel} to a scaling relation between easily measured quantities while properly accounting for both the on-orbit COS LSF as defined by \citet{kriss11} and the non-Poissonian noise characteristics of the data as described in Section~\ref{noise}.  To do so, we calculate $x_{\rm opt}$ and $f_c(x_{\rm opt})$ at 13 wavelengths and 15 $b$-values spanning the ranges $\lambda=1150$--1750~\AA\ and $b=1$--1000~\kms, respectively.  We parametrize both $x_{\rm opt}$ and $f_c(x_{\rm opt})$ as functions of the Doppler line width, $x_b$, or the number of pixels required to span a Doppler parameter, $b$, at wavelength $\lambda$:
\begin{equation}
x_b = \frac{b}{c} \; \frac{\lambda}{\Delta\lambda},
\label{eqn:dx}
\end{equation}
where $c$ is the speed of light and $\Delta\lambda$ is the dispersion of the spectrum.  This width and the Gaussian FWHM listed in Table~\ref{tab:resolved} are straightforwardly related by
\begin{equation}
\mbox{Gaussian FWHM} = 2\sqrt{\ln{2}}\,x_b \approx 1.665\,x_b.
\label{eqn:fwhm}
\end{equation}

\begin{figure}
\epsscale{1.00}
\centering \plotone{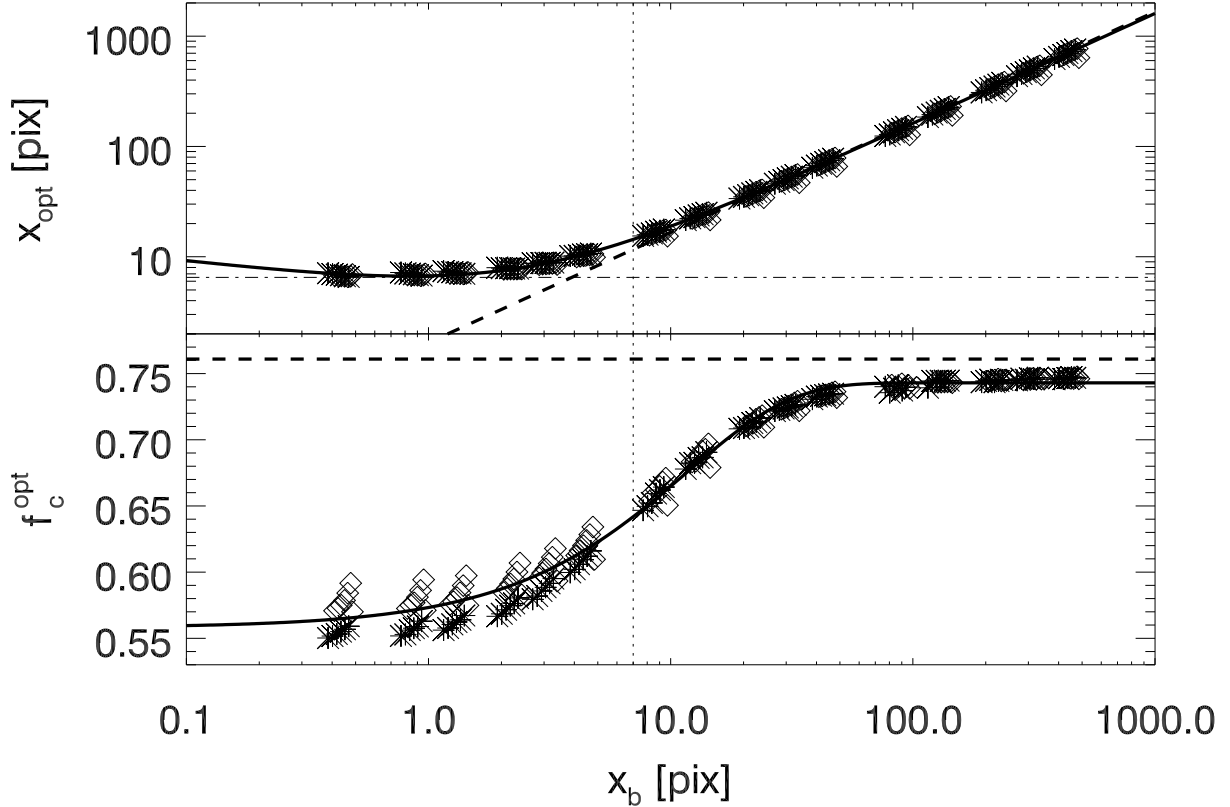}
\caption{The optimal integration width, $x_{\rm opt}$ (top panel), and the fractional area of an absorption line contained within the optimal integration width, $f_c^{\rm opt}$ (bottom panel), as a function of Doppler line width, $x_b$, evaluated for coadded COS FUV data.  Simulated G130M data are shown with asterisks and G160M data with diamonds.  The solid lines show our best fits to the data (Eq.~\ref{eqn:xopt} and \ref{eqn:fc}; Table~\ref{tab:coeff}), which hold over the entire range of $x_b$ studied although the data for $f_c^{\rm opt}$ show larger scatter when $x_b < 10$~pixels.  The dashed lines show the Gaussian relations assumed in \citet{danforth11}.  The dotted vertical line shows the approximate FWHM of the COS on-orbit LSF (see Table~\ref{tab:ewlim}), which roughly demarcates unresolved and resolved absorption lines.  The dot-dashed horizontal line in the top panel shows our recommended value of $x_{\rm opt} = 6.5$~pixels for $x_b < 1$~pixel (see text).
\label{fig:fit_coadd}}
\end{figure}

\begin{figure}
\epsscale{1.00}
\centering \plotone{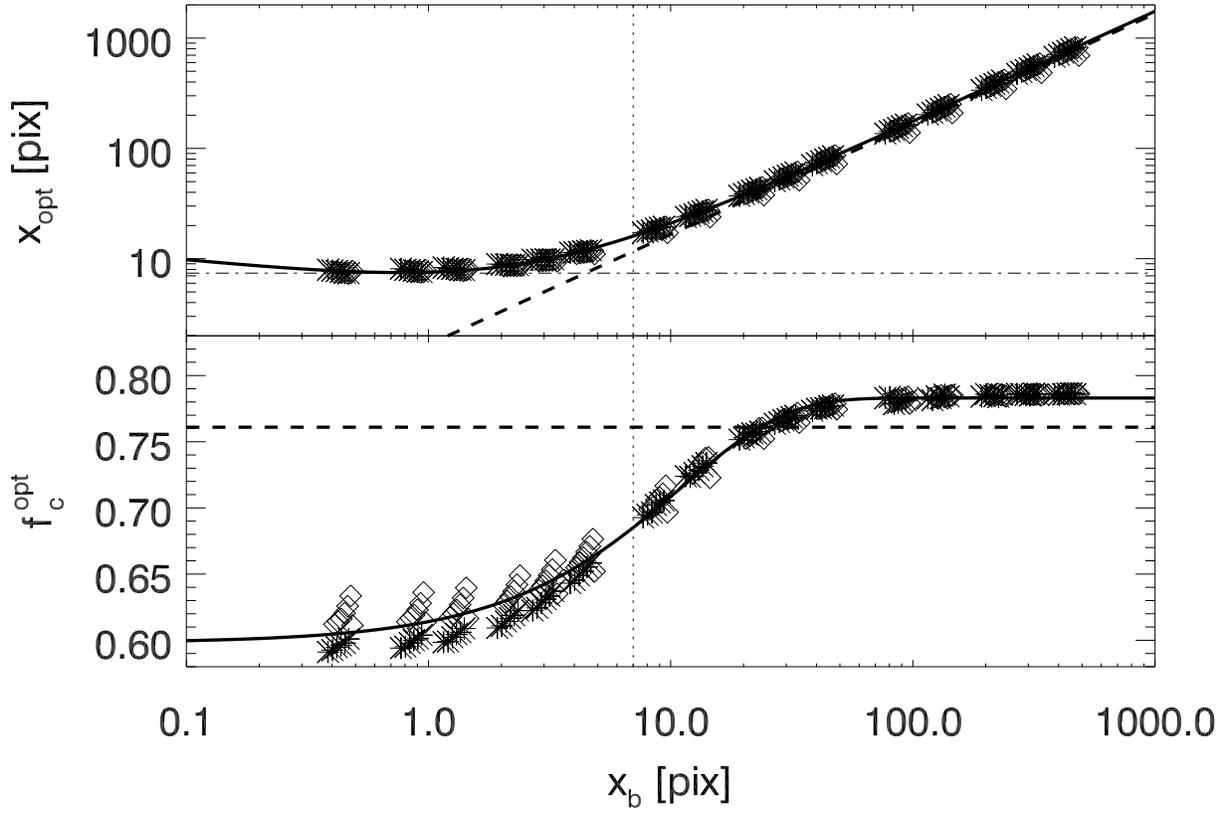}
\caption{Same as Figure~\ref{fig:fit_coadd}, but evaluated for individual COS FUV exposures.  The dot-dashed horizontal line in the top panel shows our recommended value of $x_{\rm opt} = 7.4$~pixels for $x_b < 1$~pixel (see text).
\label{fig:fit_indiv}}
\end{figure}

\begin{figure}
\epsscale{1.00}
\centering \plotone{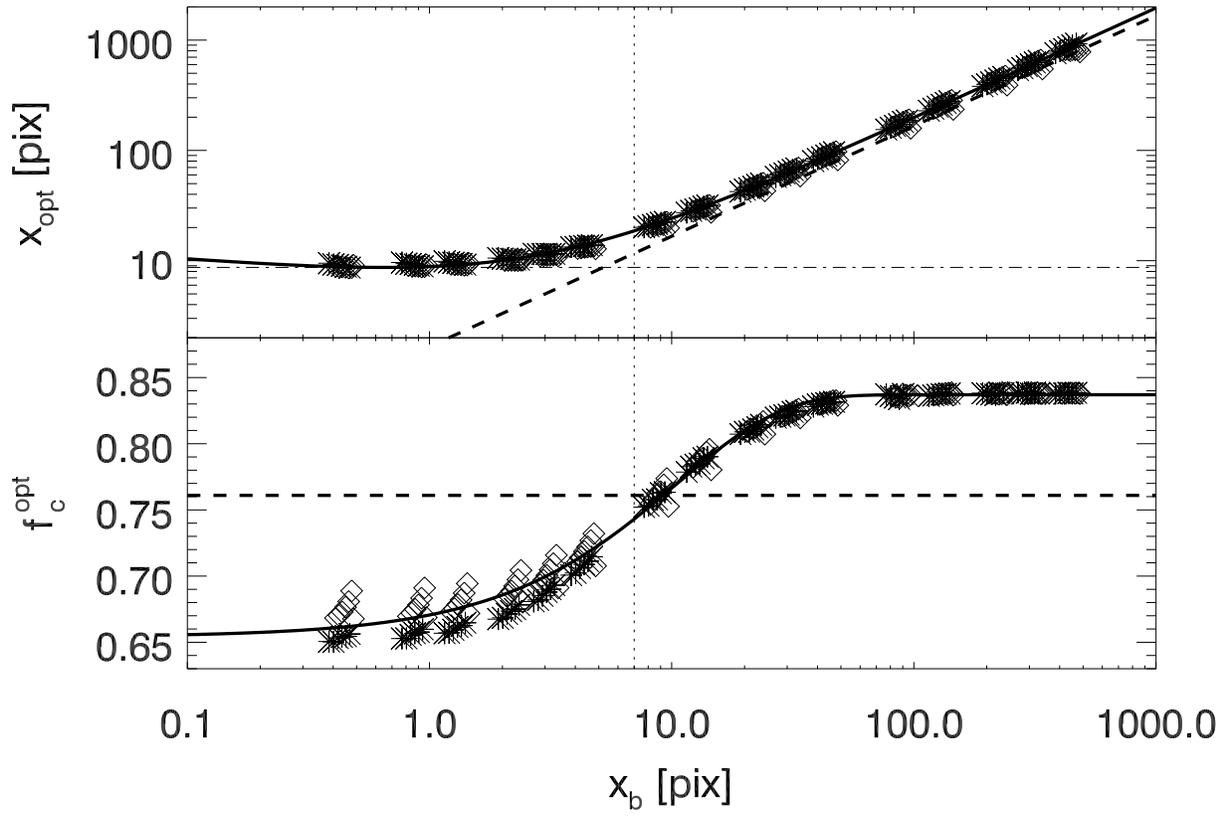}
\caption{Same as Figure~\ref{fig:fit_coadd}, but evaluated for purely Poissonian data.  The dot-dashed horizontal line in the top panel shows our recommended value of $x_{\rm opt} = 8.7$~pixels for $x_b < 1$~pixel (see text).
\label{fig:fit_poisson}}
\end{figure}

In \citet{danforth11}, we assumed that the optimal integration width equalled the Gaussian FWHM (Eq.~\ref{eqn:fwhm}), but it's clear from Table~\ref{tab:resolved} that the relationship is more complicated.  We find that $x_{\rm opt}$ is well fit by the function
\begin{equation}
x_{\rm opt}(x_b) = a_1\,x_b + b_1\,x_b^{-c_1},
\label{eqn:xopt}
\end{equation}
where the best-fit values of the coefficients $a_1$--$c_1$ for Poissonian data, individual COS exposures, and coadded COS data are listed in Table~\ref{tab:coeff}. The top panels of Figures~\ref{fig:fit_coadd}--\ref{fig:fit_poisson} show our fits to $x_{\rm opt}(x_b)$ (solid line) and the Gaussian approximation used by \citet{danforth11} (dashed line) plotted over the values derived from the LSF model of \citet{kriss11}.  The Gaussian approximation is nearly identical to our fits for $x_b \gtrsim 40$~pixels ($b \gtrsim 100$~\kms), but severely underestimates $x_{\rm opt}$ for $x_b \lesssim 10$~pixels ($b \lesssim 25$~\kms).  For very small values of the Doppler width ($x_b \lesssim 1$~pixel), Equation~\ref{eqn:xopt} overestimates the value of $x_{\rm opt}$; for these small values of $x_b$, we recommend using $x_{\rm opt} = 6.5$~pixels for coadded data, $x_{\rm opt} = 7.4$~pixels for individual exposures, and $x_{\rm opt} = 8.7$~pixels for Poissonian data, as shown by the dot-dashed horizontal lines in Figures~\ref{fig:fit_coadd}--\ref{fig:fit_poisson}.

The bottom panels of Figures~\ref{fig:fit_coadd}--\ref{fig:fit_poisson} show that $f_c(x_{\rm opt})$ asymptotically approaches values that are slightly different than the constant value that we assumed in \citet{danforth11} (dashed line). We parameterize $f_c^{\rm opt}(x_b) \equiv f_c(x_{\rm opt},\lambda,b)$ with the function
\begin{equation}
f_c^{\rm opt}(x_b) =  a_2 - b_2\,e^{-x_b/c_2},
\label{eqn:fc}
\end{equation}
which is shown by the solid lines in the bottom panels of Figures~\ref{fig:fit_coadd}--\ref{fig:fit_poisson}.  For $x_b \geq 10$~pixels, G130M data (shown with asterisks in Fig.~\ref{fig:fit_coadd}) and G160M data (diamonds) agree very well, but there is a larger amount of scatter at smaller values of $x_b$, which correspond to $b<20$~\kms\ where absorption lines are becoming increasingly unresolved.  Despite this scatter, Equation~\ref{eqn:fc} does a reasonable job of modelling $f_c^{\rm opt}(x_b)$ for all values of $x_b$ studied, particularly when compared to assuming a constant value.  The best-fit values of the coefficients $a_2$--$c_2$ for Poissonian data, individual COS exposures, and coadded COS data are listed in Table~\ref{tab:coeff}.

Substituting Equations~\ref{eqn:xopt} and \ref{eqn:fc} into Equation~\ref{eqn:siglevel} and defining $N_{\sigma}^{\rm opt}(x_b) \equiv N_{\sigma}(x_{\rm opt},\lambda,b)$, we find
\begin{align}
\label{eqn:sigapprox}
N_{\sigma}^{\rm opt}(x_b) &= {\rm (S/N)_{opt}} \, \frac{\eqw}{\Delta\lambda} \; \frac{f_c^{\rm opt}}{x_{\rm opt}} \\
                      &= {\rm (S/N)}_1 \, \frac{\eqw}{\Delta\lambda} \; \frac{1}{w(x_{\rm opt})}, \nonumber
\end{align}
where ${\rm (S/N)_{opt}}$ is the signal-to-noise measured at a resolution of $\lambda_{\rm opt} = x_{\rm opt} \times \Delta\lambda$.  We have tested the accuracy of Equation~\ref{eqn:sigapprox} over the entire COS G130M + G160M bandpass (1150--1750~\AA) at $b$-values of 1--1000~\kms\ and find that it predicts significance levels that deviate from the values derived from the LSF model of \citet{kriss11} by $\leq10$\% over the entire parameter space, and $\leq5$\% when $b>10$~\kms.

\begin{figure}
\epsscale{1.00}
\centering \plotone{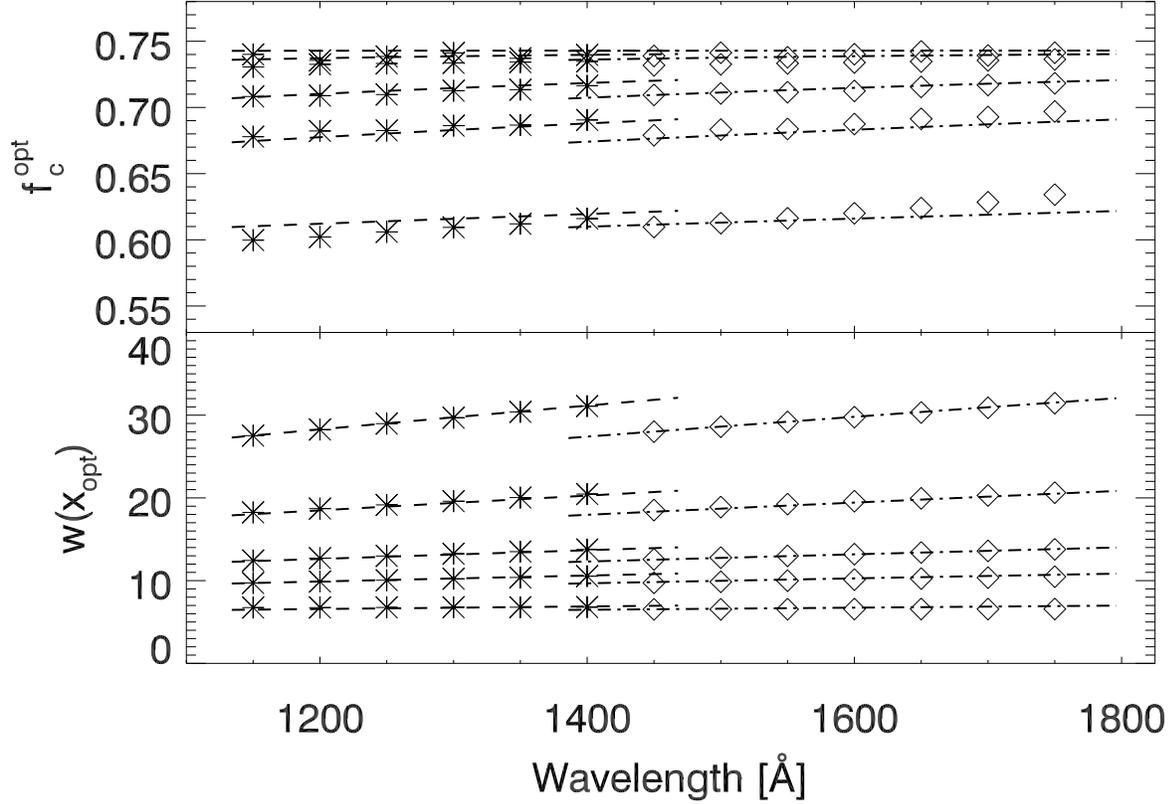}
\caption{The dependence of $f_c^{\rm opt}$ (top panel) and $w(x_{\rm opt})$ (bottom panel) on wavelength for coadded data evaluated at Doppler parameters (from top to bottom in both panels) of $b=200$, 100, 50, 30, and 10~\kms.  G130M data are shown with asterisks and G160M data with diamonds.  The values predicted by our empirical scaling relations (Equations~\ref{eqn:xopt} and \ref{eqn:fc}) are indicated by a dashed line over the G130M wavelength range and a dot-dashed line over the G160M wavelength range.
\label{fig:fw}}
\end{figure}

While it is convenient to parameterize $x_{\rm opt}$ and $f_c^{\rm opt}$, and therefore $w(x_{\rm opt})$, as a function of the single parameter $x_b$, it is still useful to examine how wavelength and Doppler parameter affect these quantities individually.  Figure~\ref{fig:fw} shows how $f_c^{\rm opt}$ and $w(x_{\rm opt})$ vary as a function of wavelength for coadded data evaluated at Doppler parameters of $b=10$, 30, 50, 100, and 200~\kms.  Both quantities have a mild wavelength dependence but vary substantially with Doppler parameter.  For modest values of the Doppler parameter ($10~\kms < b \lesssim 50~\kms$), $f_c^{\rm opt}$ increases quickly as the $b$-value increases before saturating at $f_c^{\rm opt} \approx 0.75$ (see Fig.~\ref{fig:fit_coadd} and Table~\ref{tab:coeff}).  Interestingly, $w(x_{\rm opt})$ shows the opposite behavior --- slow change for $b\lesssim 50$~\kms\ and a rapid increase for higher values.  We only present the explicit wavelength dependence of $f_c^{\rm opt}$ and $w(x_{\rm opt})$ for coadded data because we expect it will be the most applicable for observers, but the analogous behavior of individual exposures and Poissonian data can be straightforwardly derived from Equations~\ref{eqn:eta}--\ref{eqn:eta2} and \ref{eqn:dx}--\ref{eqn:fc} using the appropriate values from Table~\ref{tab:coeff}.

To facilitate the proper calculation of significance levels and equivalent width limits for coadded COS G130M and G160M data, we have written two short IDL functions that evaluate Equations~\ref{eqn:dx}--\ref{eqn:sigapprox}.  The first, \textsc{cos\_siglevel}, returns the significance level of an absorption feature in coadded data given an observed equivalent width, observed wavelength, estimated Doppler parameter, and either ${\rm (S/N)_{opt}}$ or ${\rm (S/N)}_1$ as inputs.  The second, \textsc{cos\_ewlim}, returns a limiting equivalent width for coadded data given a significance level, observed wavelength, estimated Doppler parameter, and either ${\rm (S/N)_{opt}}$ or ${\rm (S/N)}_1$ as inputs.  These functions are available on the COS Tools website at \url{http://casa.colorado.edu/$\sim$danforth/costools.html}.

\subsection{The Benefits of Coaddition}
\label{scaling:coadd}

In the previous Sections we have shown that the presence of non-Poissonian noise has an appreciable effect on the significance of absorption features detected in COS FUV data and that the non-Poissonian noise is more severe for coadded data than individual exposures.  These results may call into question the benefits of coadding COS data in the first place.  We address these questions here.

One obvious benefit to obtaining data at multiple FP-POS positions and/or central wavelength settings and coadding to obtain the final data product is that detector artifacts, such as gain sag regions \citep{sahnow11}, that are not properly removed by standard \textsc{CalCOS} processing, will be present at slightly offset wavelengths in the individual exposures.  These features move around in wavelength space precisely because they are fixed in detector (i.e., pixel) space, allowing for straightforward identification and removal, if necessary.

To assess the more subtle tradeoffs of coadding data, we compare the noise properties of coadded data and individual exposures by rearranging Equation~\ref{eqn:sigapprox} to find
\begin{equation}
\frac{w(x_{\rm opt}^{\rm coadd})}{w(x_{\rm opt}^{\rm indiv})} = \frac{{\rm (S/N)}_1^{\rm coadd}}{{\rm (S/N)}_1^{\rm indiv}} \, \frac{N_{\sigma}^{\rm indiv}}{N_{\sigma}^{\rm coadd}} \, \frac{\eqw^{\rm coadd}}{\eqw^{\rm indiv}} \, \frac{\Delta\lambda^{\rm indiv}}{\Delta\lambda^{\rm coadd}},
\label{eqn:wratio}
\end{equation}
where $w(x_{\rm opt}^{\rm coadd})$ and $w(x_{\rm opt}^{\rm indiv})$ are $w(x_{\rm opt})$ evaluated for coadded data and individual exposures, respectively, and the other terms are defined analogously.  This ratio is plotted as a function of $x_b$ in Figure~\ref{fig:wratio}, which shows that $w(x_{\rm opt}^{\rm coadd})$ is approximately 6\% larger than $w(x_{\rm opt}^{\rm indiv})$ for unresolved lines, and their ratio increases as $x_b^{0.053}$ (dashed line) for resolved lines ($x_b\gtrsim10$~pixels).

Equation~\ref{eqn:wratio} can be interpreted in several useful ways.  If one assumes that the signal-to-noise, significance level, and dispersion of coadded data and individual exposures are the same, then Equation~\ref{eqn:wratio} quantifies how much larger limiting equivalent widths for coadded data are than limiting equivalent widths for individual exposures.  Similarly, if one assumes that the signal-to-noise, equivalent width, and dispersion of coadded data and individual exposures are the same, then Equation~\ref{eqn:wratio} quantifies how much larger the significance level for individual exposures are than the significance level for coadded data.  Finally, and most germane to this discussion, if one assumes that the dispersion of coadded data and individual exposures are the same, then Equation~\ref{eqn:wratio} quantifies the improvement in signal-to-noise ratio required for an absorption feature of a given equivalent width to have the same significance level in coadded data and individual exposures.

If two exposures of equal exposure time are coadded, then one would (na\"ively) expect that ${\rm (S/N)}_1^{\rm coadd} = \sqrt{2}\,{\rm (S/N)}_1^{\rm indiv}$, in which case the increased signal-to-noise achieved through coaddition compensates for the increased non-Poissonian noise introduced by the coaddition process when $x_b \lesssim 700$~pixels ($b \lesssim 1500$~\kms).  We have made no efforts to test how signal-to-noise increases as a function of exposure time for COS data, but we note that even if coadding two equal-length exposures only increases the signal-to-noise per pixel by a factor of $2^{0.37} \approx 1.29 $ (chosen to match the exponent $\beta$ found when characterizing $\eta(x)$ for coadded data; see Table~\ref{tab:coeff}), then coadding will still increase the overall sensitivity for $x_b \lesssim 125$~pixels ($b \lesssim 300$~\kms).  

Therefore, coadding two equal-length exposures will improve the limiting equivalent width for all but the broadest absorption lines.  The limiting equivalent widths of even the broadest lines can be improved by adding more than two exposures together since we do not find any dependence of $\eta(x)$ on the number of exposures contributing to the coaddition (Section~\ref{noise}).  Since most COS FUV observations consist of four or more exposures per grating, coaddition will almost always improve the sensitivity of a dataset to weak absorption features.

This happy result is largely a consequence of the individual exposures themselves being non-Poissonian.  If we compare coadded COS data to Poissonian data we find that $w(x_{\rm opt}^{\rm coadd})$ is approximately 20\% larger than $w(x_{\rm opt}^{\rm Poisson})$ for unresolved lines, and their ratio increases as $x_b^{0.142}$ for resolved lines.  Thus, if individual COS exposures were Poissonian then coadding two equal-length exposures would only improve the limiting equivalent width for absorption lines with $b\lesssim25$~\kms, assuming ${\rm (S/N)}_1^{\rm coadd} \propto t^{1/2}$.

\section{Conclusions}
\label{conclusion}

\citet{ghavamian09} and \citet{kriss11} showed that the extended wings of the on-orbit LSFs of the COS medium resolution gratings lead to limiting equivalent widths for unresolved absorption lines that are $\sim30$--40\% higher than those derived from a Gaussian LSF, assuming Poissonian noise properties.  We have demonstrated that COS G130M and G160M data have non-Poissonian noise characteristics and that the smoothed signal-to-noise ratio, ${\rm (S/N)}_x$, is proportional to $x^{0.42}$ for individual exposures and $x^{0.37}$ for optimally coadded data (see discussion in Section~\ref{noise}), rather than $x^{1/2}$ as Poissonian statistics predict.  When this non-Poissonian noise is accounted for, the limiting equivalent width of unresolved absorption lines increases by $\sim15$--20\% over the \citet{ghavamian09} and \citet{kriss11} values.

We have also extended the equivalent width formalism of \citet{ghavamian09} to spectroscopically resolved lines and find that the non-Poissonian noise properties of COS data have an even larger effect on resolved lines than unresolved lines.  For a Doppler parameter of $b=10$~\kms, the limiting equivalent width for coadded data is 30\% larger than the corresponding Poissonian value; this percentage grows to 90\% larger than the Poissonian value for $b=200$~\kms.  This effect is caused by the smoothed signal-to-noise differing more and more from the Poissonian value as the integration width increases, and emphasizes the importance of properly accounting for the non-Poissonian noise properties of COS G130M and G160M data.

To that end, we have derived empirical scaling relations (Eqs.~\ref{eqn:dx}--\ref{eqn:sigapprox}) that allow easy calculation of the optimal significance of a detected absorption feature, $N_{\sigma}^{\rm opt}$, as a function of the Doppler line width, $x_b$.  These relations have been compared to the values derived from the LSF model of \citet{kriss11} in the wavelength range $\lambda=1150$--1750~\AA\ and Doppler parameter range $b=1$--1000~\kms\ and found to differ by $\leq10$\% over the whole parameter range, and $\leq5$\% when $b>10$~\kms.  IDL functions that use our parameterizations to determine the significance level of an observed feature or set an equivalent width limit on a nondetection have been developed and made publicly available (see Section~\ref{scaling}).

\begin{figure}
\epsscale{1.00}
\centering \plotone{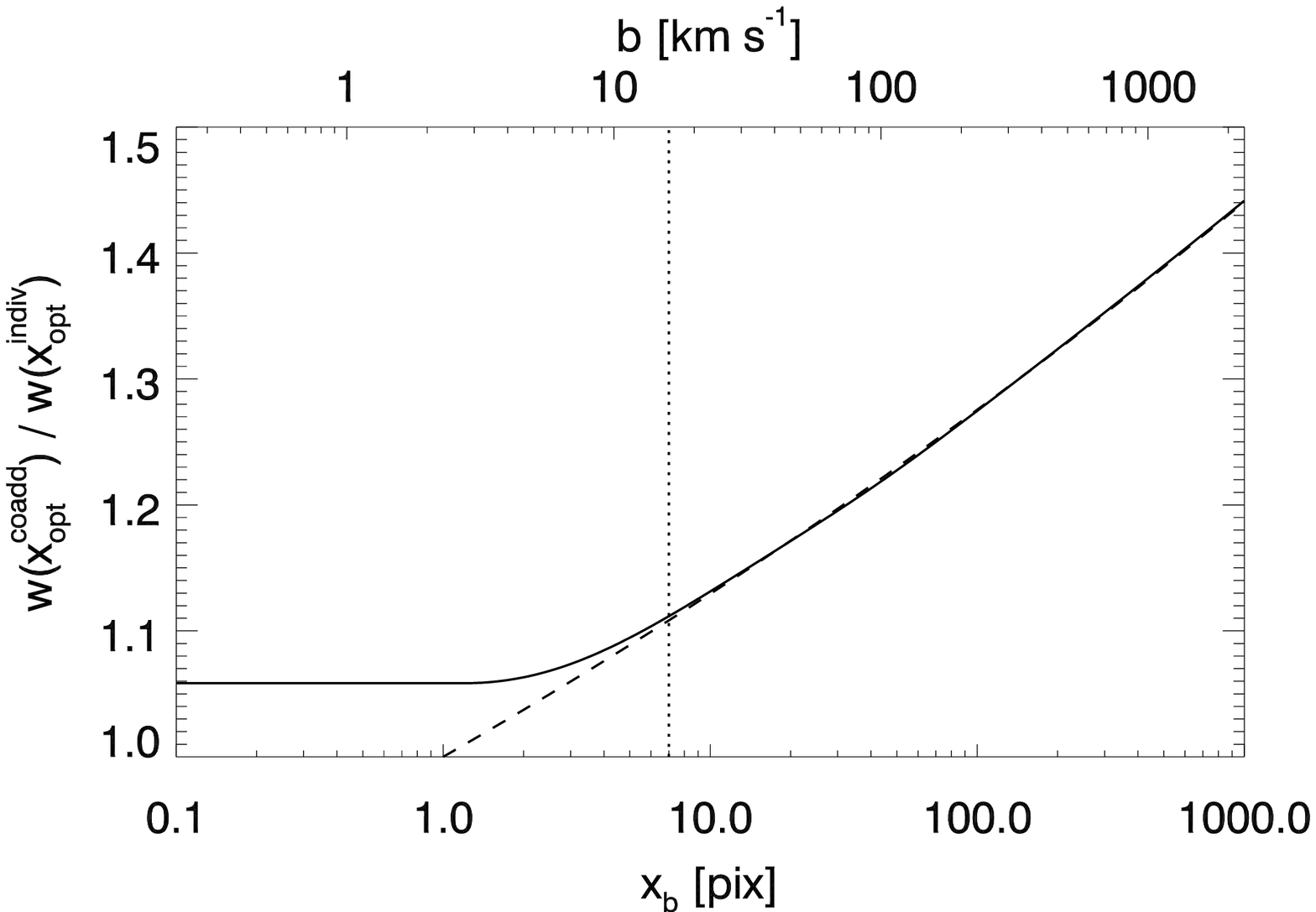}
\caption{The ratio $w(x_{\rm opt}^{\rm coadd})/w(x_{\rm opt}^{\rm indiv})$ (see Eq.~\ref{eqn:wratio}), plotted both as a function of Doppler line width, $x_b$ (bottom axis), and as an approximate function of Doppler parameter, $b$ (top axis, for which the labelled values are only strictly valid when $\lambda=1450$~\AA).  The dotted vertical line shows the approximate FWHM of the COS on-orbit LSF (see Table~\ref{tab:ewlim}), which roughly demarcates unresolved and resolved absorption lines.  The dashed line shows the function $x_b^{0.053}$, which is an excellent fit to the plotted ratio for resolved lines.
\label{fig:wratio}}
\end{figure}

While the non-Poissonian noise in coadded data is more severe than in individual exposures, the fact that the individual exposures are themselves non-Poissonian means that carefully coadding data can lead to significant reductions of the limiting equivalent width (i.e., improvements in sensitivity).  Taking the non-Poissonian noise properties of both individual exposures and coadded data into account, we find that the signal-to-noise per pixel of coadded data need only increase by 6\% over the signal-to-noise in individual exposures to reduce the limiting equivalent width of unresolved lines.  For resolved lines, the necessary improvement in signal-to-noise scales approximately as $x_b^{0.053}$ for $x_b \gtrsim 10$~pixels (see Fig.~\ref{fig:wratio}).  Thus, a signal-to-noise improvement of $\lesssim25$\% is required for coadded data to have higher sensitivity than individual exposures for lines with $b\lesssim150$~\kms.

\acknowledgments
This work was supported by NASA grants NNX08AC146 and NAS5-98043 to the University of Colorado at Boulder for the \hst/COS project.  We also thank T.~Tripp and J.~Tumlinson for helpful discussions, and D.~Syphers for detailed comments on the manuscript.

{\it Facilities:} \facility{HST (COS)}


\clearpage
\begin{deluxetable}{lcccc}

\tablecolumns{5}
\tablewidth{0pt}

\tablecaption{Best-Fit Coefficients
\label{tab:coeff}}

\tablehead{ & \colhead{Poissonian} & \colhead{Individual} & \colhead{Coadded} & \\ \colhead{Parameter} & \colhead{Data} & \colhead{Exposures} & \colhead{Data} & \colhead{Eqn}}

\startdata
$\eta_0$\dotfill &  0              & $0.08\pm0.01$   & $0.15\pm0.02$   & (4) \\
$\beta$\dotfill  & 1/2             & $0.42\pm0.01$   & $0.37\pm0.01$   & (4) \\
$a_1$\dotfill    & $1.954\pm0.003$ & $1.748\pm0.002$ & $1.605\pm0.002$ & (9) \\
$b_1$\dotfill    & $6.9\pm0.1$     & $5.8\pm0.1$     & $5.1\pm0.1$     & (9) \\
$c_1$\dotfill    & $0.17\pm0.01$   & $0.22\pm0.01$   & $0.25\pm0.01$   & (9) \\
$a_2$\dotfill    & $0.837\pm0.001$ & $0.783\pm0.001$ & $0.743\pm0.001$ & (10) \\
$b_2$\dotfill    & $0.183\pm0.005$ & $0.185\pm0.005$ & $0.185\pm0.004$ & (10) \\
$c_2$\dotfill    & $10.5\pm0.4$    & $11.0\pm0.4$    & $11.6\pm0.4$    & (10)
\enddata

\end{deluxetable}

\begin{deluxetable}{lcccccccccccccc}

\rotate
\tabletypesize{\footnotesize}

\tablecolumns{15}
\tablewidth{0pt}

\tablecaption{Limiting Equivalent Widths for Unresolved Absorption Lines
\label{tab:ewlim}}

\tablehead{ & & \multicolumn{3}{c}{Ghavamian LSF (Poisson)} && \multicolumn{3}{c}{Kriss LSF (Poisson)} && \multicolumn{2}{c}{Kriss LSF (Indiv. Exp.)} && \multicolumn{2}{c}{Kriss LSF (Coadded)} \\  \cline{3-5} \cline{7-9} \cline{11-12} \cline{14-15} \colhead{Grating} & \colhead{$\lambda$} & \colhead{FWHM} & \colhead{$x_{\rm opt}$} & \colhead{$\ewlim(3\sigma)$} && \colhead{FWHM} & \colhead{$x_{\rm opt}$} & \colhead{$\ewlim(3\sigma)$} && \colhead{$x_{\rm opt}$} & \colhead{$\ewlim(3\sigma)$} && \colhead{$x_{\rm opt}$} & \colhead{$\ewlim(3\sigma)$} \\ & \colhead{(\AA)} & \colhead{(pixels)} & \colhead{(pixels)} & \colhead{(m\AA)} && \colhead{(pixels)} & \colhead{(pixels)} & \colhead{(m\AA)} && \colhead{(pixels)} & \colhead{(m\AA)} && \colhead{(pixels)} & \colhead{(m\AA)}}

\startdata
G130M & Gauss & 6.5 & 7.6 & ~9.9 && 6.5 & 7.6 & ~9.9 && 6.6 & 11.1 && 5.8 & 11.7 \\
G130M & 1150  & 7.5 & 9.2 & 13.7 && 7.4 & 9.2 & 14.1 && 7.6 & 16.2 && 6.8 & 17.2 \\
G130M & 1200  & 7.3 & 9.0 & 13.5 && 7.3 & 9.0 & 13.9 && 7.6 & 15.9 && 6.6 & 17.0 \\
G130M & 1250  & 7.3 & 9.0 & 13.4 && 7.2 & 9.0 & 13.8 && 7.4 & 15.8 && 6.6 & 16.8 \\
G130M & 1300  & 7.2 & 8.8 & 13.3 && 7.1 & 8.8 & 13.6 && 7.4 & 15.6 && 6.6 & 16.6 \\
G130M & 1350  & 7.0 & 8.6 & 13.1 && 7.0 & 8.6 & 13.5 && 7.2 & 15.4 && 6.4 & 16.3 \\
G130M & 1400  & 7.0 & 8.4 & 13.0 && 6.9 & 8.4 & 13.3 && 7.2 & 15.2 && 6.2 & 16.1 \\
\\
G160M & Gauss & 6.5 & 7.5 & 12.1 && 6.5 & 7.5 & 12.1 && 6.5 & 13.7 && 5.9 & 14.4 \\
G160M & 1450  & 7.3 & 9.0 & 16.1 && 7.4 & 9.0 & 16.5 && 7.5 & 18.9 && 6.7 & 20.1 \\
G160M & 1500  & 7.1 & 8.8 & 15.9 && 7.2 & 8.7 & 16.3 && 7.4 & 18.6 && 6.5 & 19.8 \\
G160M & 1550  & 7.1 & 8.7 & 15.9 && 7.1 & 8.7 & 16.2 && 7.4 & 18.5 && 6.5 & 19.7 \\
G160M & 1600  & 7.1 & 8.7 & 15.8 && 7.0 & 8.7 & 16.1 && 7.4 & 18.3 && 6.5 & 19.5 \\
G160M & 1650  & 7.0 & 8.5 & 15.6 && 7.0 & 8.5 & 15.9 && 7.2 & 18.1 && 6.4 & 19.3 \\
G160M & 1700  & 6.9 & 8.5 & 15.4 && 7.0 & 8.3 & 15.7 && 7.2 & 17.9 && 6.4 & 19.0 \\
G160M & 1750  & 7.0 & 8.5 & 15.3 && 7.0 & 8.5 & 15.6 && 7.2 & 17.7 && 6.4 & 18.8
\enddata

\tablecomments{All equivalent width limits assume ${\rm S/N} = 10$ per pixel. The two line spread functions are detailed in \citet{ghavamian09} and \citet{kriss11}.}

\end{deluxetable}

\begin{deluxetable}{cccccccccccccc}

\rotate
\tabletypesize{\footnotesize}

\tablecolumns{14}
\tablewidth{0pt}

\tablecaption{Limiting Equivalent Widths for Resolved Absorption Features
\label{tab:resolved}}

\tablehead{ & & & \multicolumn{3}{c}{Poisson} && \multicolumn{3}{c}{Indiv. Exp.} && \multicolumn{3}{c}{Coadded} \\ \cline{4-6} \cline{8-10} \cline{12-14} & \colhead{Gaussian} & \colhead{Convolved} \\ \colhead{$b$} & \colhead{FWHM} & \colhead{FWHM} & \colhead{$x_{\rm opt}$} & \colhead{$f_c(x_{\rm opt})$} & \colhead{$\ewlim(3\sigma)$} && \colhead{$x_{\rm opt}$} & \colhead{$f_c(x_{\rm opt})$} & \colhead{$\ewlim(3\sigma)$} && \colhead{$x_{\rm opt}$} & \colhead{$f_c(x_{\rm opt})$} & \colhead{$\ewlim(3\sigma)$} \\ \colhead{(\kms)} & \colhead{(pixels)} & \colhead{(pixels)} & \colhead{(pixels)} &  & \colhead{(m\AA)} && \colhead{(pixels)} &  & \colhead{(m\AA)} && \colhead{(pixels)} &  & \colhead{(m\AA)}}

\startdata
\multicolumn{14}{c}{{\em G130M} ($\lambda = 1250$~\AA)} \\
\tableline
~10 & ~~7.0 & ~10.9 & ~13.2 & 0.706 & 15.5 && ~11.4 & 0.649 & 18.4 && ~10.0 & 0.605 & 20.1 \\
~20 & ~13.9 & ~17.5 & ~21.1 & 0.758 & 18.1 && ~18.1 & 0.696 & 22.5 && ~16.3 & 0.652 & 25.2 \\
~30 & ~20.9 & ~24.3 & ~29.1 & 0.784 & 20.6 && ~25.3 & 0.726 & 26.3 && ~22.9 & 0.683 & 30.1 \\
~40 & ~27.9 & ~31.2 & ~37.3 & 0.801 & 22.8 && ~32.5 & 0.743 & 29.8 && ~29.5 & 0.699 & 34.6 \\
~50 & ~34.8 & ~38.1 & ~45.3 & 0.809 & 24.9 && ~39.9 & 0.755 & 33.1 && ~36.1 & 0.710 & 38.8 \\
~75 & ~52.2 & ~55.3 & ~66.0 & 0.824 & 29.5 && ~58.0 & 0.768 & 40.5 && ~53.0 & 0.726 & 48.5 \\
100 & ~69.6 & ~72.6 & ~86.5 & 0.830 & 33.5 && ~76.2 & 0.775 & 47.0 && ~69.6 & 0.733 & 57.3 \\
150 & 104.5 & 107.1 & 127.4 & 0.835 & 40.4 && 112.5 & 0.781 & 58.6 && 103.1 & 0.740 & 73.0 \\
200 & 139.3 & 141.6 & 168.3 & 0.837 & 46.4 && 149.1 & 0.784 & 68.9 && 136.4 & 0.742 & 87.0 \\
\\
\multicolumn{14}{c}{{\em G160M} ($\lambda = 1550$~\AA)} \\
\tableline
~10 & ~~7.0 & ~10.7 & ~12.9 & 0.715 & 18.5 && ~11.1 & 0.659 & 21.9 && ~10.0 & 0.616 & ~23.9 \\
~20 & ~14.1 & ~17.2 & ~20.6 & 0.758 & 22.0 && ~17.8 & 0.700 & 27.2 && ~16.2 & 0.660 & ~30.5 \\
~30 & ~21.1 & ~24.1 & ~29.0 & 0.786 & 25.1 && ~25.2 & 0.729 & 32.1 && ~22.7 & 0.684 & ~36.7 \\
~40 & ~28.2 & ~31.2 & ~37.3 & 0.801 & 28.0 && ~32.5 & 0.744 & 36.5 && ~29.6 & 0.702 & ~42.4 \\
~50 & ~35.2 & ~38.1 & ~45.6 & 0.811 & 30.6 && ~39.9 & 0.754 & 40.6 && ~36.3 & 0.712 & ~47.7 \\
~75 & ~52.8 & ~55.7 & ~66.4 & 0.824 & 36.3 && ~58.4 & 0.768 & 49.8 && ~53.3 & 0.727 & ~59.8 \\
100 & ~70.4 & ~73.2 & ~87.2 & 0.830 & 41.3 && ~76.9 & 0.776 & 58.0 && ~70.2 & 0.733 & ~70.6 \\
150 & 105.6 & 108.1 & 128.5 & 0.835 & 49.8 && 113.7 & 0.781 & 72.3 && 104.0 & 0.740 & ~90.1 \\
200 & 140.8 & 143.0 & 170.1 & 0.837 & 57.1 && 150.6 & 0.784 & 85.0 && 137.9 & 0.743 & 107.5 
\enddata

\tablecomments{All equivalent width limits assume ${\rm S/N} = 10$ per pixel.  All convolutions use the COS line spread function detailed in \citet{kriss11}.}

\end{deluxetable}

\clearpage

\begin{thebibliography}

\bibitem[Ake et~al.(2010)]{ake10} Ake, T. B., Massa, D., B\'eland, S., et~al. 2010, in The 2010 STScI Calibration Workshop, eds. S. Deustua \& C. Oliveira (Baltimore: STScI), p. 23

\bibitem[Cardelli \& Ebbets(1993)]{cardelli93} Cardelli, J. A. \& Ebbets, D. C. 1993, in Calibrating Hubble Space Telescope, eds. J .C. Blades and S. J. Osmer (Baltimore: STScI), p. 322

\bibitem[Danforth et~al.(2010a)]{danforth10a} Danforth, C. W., Keeney, B. A., Stocke, J. T., Shull, J. M., \& Yao, Y. 2010, \apj, 720, 976

\bibitem[Danforth et~al.(2010b)Danforth, Stocke, \& Shull]{danforth10b} Danforth, C. W., Stocke, J. T., \& Shull, J. M. 2010, \apj, 710, 613

\bibitem[Danforth et~al.(2011)]{danforth11} Danforth, C. W., Stocke, J. T., Keeney, B. A., et~al. 2011, \apj, 743, 18

\bibitem[Dixon et~al.(2011)]{dixon11} Dixon, W. V., et~al. 2011, Cosmic Origins Spectrograph Instrument Handbook, Version 4.0 (Baltimore: STScI)

\bibitem[Ghavamian et~al.(2009)]{ghavamian09} Ghavamian, P., Aloisi, A., Lennon, D., et~al. 2009, COS Instrument Science Report 2009-01(v1), Preliminary Characterization of the Post-Launch Line Spread Function of COS (Baltimore: STScI)

\bibitem[Green et~al.(2012)]{green12} Green, J. C., Froning, C. S., Osterman, S., et~al. 2012, \apj, 744, 60

\bibitem[Keeney et~al.(2012)]{keeney12} Keeney, B. A., Stocke, J. T., Rosenberg, J. L., et~al. 2012, \apj, submitted

\bibitem[Kriss(2011)]{kriss11} Kriss, G. A. 2011, COS Instrument Science Report 2011-01(v1), Improved Medium Resolution Line Spread Functions for COS FUV Spectra (Baltimore: STScI)

\bibitem[Lambert et~al.(1994)]{lambert94} Lambert, D. L., Sheffer, Y., Gilliland, R. L., \& Federman, S. R. 1994, \apj, 420, 756

\bibitem[Narayanan et~al.(2011)]{narayanan11} Narayanan, A., Savage, B. D., Wakker, B. P., et~al. 2011, \apj, 730, 15

\bibitem[Osterman et~al.(2011)]{osterman11} Osterman, S., Green, J., Froning, C., et~al. 2011, \apss, 335, 257

\bibitem[Sahnow et~al.(2011)]{sahnow11} Sahnow, D. J., Oliveira, C., Aloisi, A., et~al. 2011, in Proc. SPIE 8145, UV, X-ray, and Gamma-Ray Space Instrumentation for Astronomy XVII., ed. L. Tsakalakos (AIP), 81450Q

\bibitem[Savage et~al.(2010)]{savage10} Savage, B. D., Narayanan, A., Wakker, B. P., et~al. 2010, \apj, 721, 960

\bibitem[Savage et~al.(2011)]{savage11} Savage, B. D., Narayanan, A., Lehner, N., \& Wakker, B. P. 2011, \apj, 731, 14

\bibitem[Syphers et~al.(2012)]{syphers12} Syphers, D., Anderson, S. F., Zheng, W., et~al. 2012, \aj, 143, 100

\end{thebibliography}
\end{document}